\documentclass[11pt,aps,nofootinbib,
floatfix,superscriptaddress]{revtex4}
\usepackage{graphicx}
\usepackage[mathscr]{eucal}
\usepackage{hyperref}
\begin{document}

\title{Quark-diquark structure and masses of doubly charmed baryons}

\begin{abstract}
{The finite size of doubly heavy diquark gives a positive correction to the masses of baryons calculated in the local diquark approximation. We evaluate this correction for the basic states of doubly charmed baryons to give quite accurate predictions actual for current searches of those baryons at LHCb~\cite{Aaij:2013voa}: $m[{\Xi_{cc}^{1/2}}^{+}]\approx m[{\Xi_{cc}^{1/2}}^{++}]=3615\pm 55$ MeV and 
 $m[{\Xi_{cc}^{3/2}}^{+}]\approx m[{\Xi_{cc}^{3/2}}^{++}]=3747\pm 55$ MeV.}
\end{abstract}

\author{V.V.Kiselev}
\email{Valery.Kiselev@ihep.ru; kiselev.vv@mipt.ru}

\affiliation{Department of General and Applied Physics, 
Moscow Institute of Physics and Technology (State University),
Russia, 141701, Moscow Region, Dolgoprudny, Institutsky 9} 
\affiliation{State
Research Center of the Russian Federation  ``Institute for High Energy
Physics'' of National Research Centre  ``Kurchatov Institute'',
Russia, 142281, Moscow Region, Protvino, Nauki 1}

\author{A.V.Berezhnoy}
\email{alexander.berezhnoy@cern.ch}

\affiliation{ Lomonosov Moscow State University, Skobeltsyn Institute of Nuclear Physics (SINP MSU), 1(2), Leninskie gory, GSP-1, Moscow 119991, Russian Federation}

\author{A.K.Likhoded}
\email{Anatolii.Likhoded@ihep.ru}

\affiliation{Department of General and Applied Physics, 
Moscow Institute of Physics and Technology (State University),
Russia, 141701, Moscow Region, Dolgoprudny, Institutsky 9}
\affiliation{State
Research Center of the Russian Federation  ``Institute for High Energy
Physics'' of National Research Centre  ``Kurchatov Institute'',
Russia, 142281, Moscow Region, Protvino, Nauki 1}

\maketitle

\section{Introduction}
Solving the problem of quark-gluon confinement in QCD demands a deep understanding of mechanisms dealing with the quark-hadron duality \cite{Olive:2016xmw}. In particlular, the quark-diquark structure of baryons \cite{Richard:1992uk,Karliner:2017kfm} as well as consequent issues on the composition of exotic hadrons \cite{Karliner:2016zzc,Ali:2017jda,Berezhnoy:2011xn} can be thoroughly studied in the spectroscopy of doubly heavy baryons \cite{Fleck:1989mb,Kiselev:2001fw,Anikeev:2001rk,Gershtein:2000nx,Kiselev:2002iy}. 

In this way, to the leading order the diquark composed of two heavy quarks can be approximated as a local object, since its size can tend to zero in the limit of infinitely heavy quarks, hence, a quark-gluon string of length $r$ as modelled by the linear confining potential $V_c = \sigma\cdot r$ can be presented as a line connecting the light quark to the diquark center. The string tension is well fitted in the spectroscopy, $\sigma=0.18$ GeV$^2$. The mass of doubly heavy diquark can be accurately predicted in non-relativistic potential models adjusted by the spectra of heavy quarkonia, so that the uncertainty is limited by 30 MeV. To the other hand, the spectroscopy of heavy-light mesons allows us to calculate the masses of such the bound heavy-light system composed of local doubly heavy diquark and light quark. Therefore, with accuracy up to the isotopic splitting of about 2-3 MeV the local-diquark approximation easily predict the masses of doubly charmed baryons \cite{Gershtein:2000nx,Gershtein:1998sx}:

for the baryon spin $1/2$
$$
	m[{\Xi_{cc}^{1/2}}^{++}]\approx m[{\Xi_{cc}^{1/2}}^+]=3478\pm 30 \mbox{ MeV,}
$$

for the baryon spin $3/2$
$$
	m[{\Xi_{cc}^{3/2}}^{++}]\approx m[{\Xi_{cc}^{3/2}}^+]=3610\pm 30 \mbox{ MeV.}
$$

However, the actual sizes of doubly charmed diquarks are not negligible. As we have found in \cite{Gershtein:2000nx,Gershtein:1998sx}, the size of basic vector $1S$-diquark is approximately equal to $r_d\approx 0.6$ Fermi $\approx 3$ GeV$^{-1}$.  Therefore, the contribution of quark-gluon string into the binding energy of diquark is essential, and it was taken into account in terms of confining linear potential for the straightforward string between two heavy quarks. 
Such the straightforward string is equivalent to the local diquark approximation\footnote{Other approaches to the evaluating the masses of doubly heavy baryons are presented in \cite{Ebert:1996ec,Flynn:2003vz,Albertus:2006ya,
Brambilla:2005yk,Shah:2017liu}.}, indeed.

In order to take into account the finite size of doubly heavy diquark we have to consider the distorsion of straightforward string inside the diquark. In this respect the doubly charmed baryons provide us with a unique opportunity to research the diquark system due to the identity of two heavy flavored quarks in the diquark, since in addition to the basic state of vector heavy $1S$-diquark the vector heavy $2P$-diquark is expected to be quazi-stable because the transition of vector $2P$-diquark into vector $1S$-diquark can take place only due to the change of both the spin and orbital momentum of the diquark that is suppressed by second power of inverse heavy quark mass being much less than the radius of strong interaction. On the other hand, the sizes of these two heavy diquarks are significantly different, hence, the effects of finite size of diquarks can be discriminated.

\section{Estimates of corrections by the diquark string distorsion}
For three light quarks the string inside the baryon has got the from of symmetric three-ray star (see Fig. 1). Essentially, there is the central point of string connecting the light quark to the diquark. In that case the string length between the quarks inside the diquark is greater than the distance between those quarks. Therefore, the correction caused by the finite size of diquark is given by
\begin{equation}\label{3-star}
	\delta E_\ast = \sigma \cdot \delta r_\ast =\sigma\cdot\,r_d\cdot \frac{2-\sqrt{3}}{\sqrt{3}},
\end{equation}
i.e. the longer string gives greater mass of diquark. The correction is proportional to the size of doubly heavy diquark, so that the limit of infinitely heavy quarks is well satisfied. Numerically, the length is enlarged to 15\%.

\begin{figure}[h]
\begin{center}
\includegraphics[width=2.5cm]{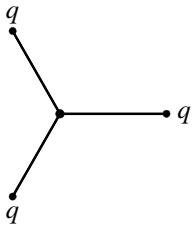}
\caption{The quark-gluon string confining three light quarks.}
\label{fig-1}
\end{center}
\end{figure}

The other string configuration presents the maximal distortion of string: the point of string connection to the diquark is posed at the sphere with the radius equal to the diquark size (see Fig. 2). Then, the maximal enhancement of string length inside the diquark gives the correction to the diquark mass equal to 
\begin{equation}\label{star-max}
	\delta E_\mathrm{max} = \sigma \cdot \delta r_\mathrm{max} =\sigma\cdot\,r_d\cdot (\sqrt{2}-1),
\end{equation}
Numerically, the length is enlarged to 41\%.

\begin{figure}[h]
\begin{center}
 \includegraphics[width=4cm]{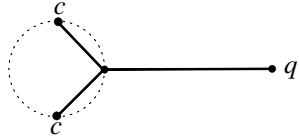}
\caption{The quark-gluon string connecting the light quark with the surface of doubly heavy diquark.}
\label{fig-2}
\end{center}
\end{figure}

Let us use the average energy shift caused by the finite size of diquark
\begin{equation}\label{ave}
	\delta E=\frac12 (\delta E_\ast+\delta E_\mathrm{max}) = \sigma \cdot r_d \cdot(0.28\pm 0.09),
\end{equation}
wherein we suppose that the average value determines the correction with the accuracy covering 69\% of interval between the boarder points.
 
We see that the correction is positive due to longer string inside the non-local diquark. The estimate has a valuable uncertainty. Nevertheless, the direct application to the doubly heavy baryons results in 
\begin{equation}\label{ave-2}
	\delta E[\Xi_{cc}]=137\pm 46 \mbox{ MeV}.
\end{equation}

Summing up uncertainties caused by the accuracy of potential models with the local diquark and that of in the energy shift due to the finite size of diquark we get estimates 

for the baryon spin $1/2$
\begin{equation}
	m[{\Xi_{cc}^{1/2}}^{++}]\approx m[{\Xi_{cc}^{1/2}}^+]=3615\pm 55 \mbox{ MeV,}
\end{equation}

for the baryon spin $3/2$
\begin{equation}
	m[{\Xi_{cc}^{3/2}}^{++}]\approx m[{\Xi_{cc}^{3/2}}^+]=3747\pm 55 \mbox{ MeV.}
\end{equation}

As we have mentioned in Introduction, the diquark composed of two identical charmed quarks possesses the quazi-stable vector doubly charmed $2P$-state with zeroth summed spin of quarks, since the transition to the vector $1S$-state with summed spin equal to 1 takes place with the change of both the summed spin and relative orbital momentum, $\Delta S=\Delta L=1$.
The size of such diquark is essentially greater than the size of basic vector diquark: $r_{d(2P)}\approx 0.9$ Fermi \cite{Gershtein:2000nx,Gershtein:1998sx}, hence, the correction exceeds the value in (\ref{ave-2}) by 33 \%:
\begin{equation}\label{ave-3}
	\delta E[\Xi_{cc(2P)}]=183\pm 59 \mbox{ MeV}.
\end{equation}
Thus, for the quazi-stable doubly charmed baryons we get estimates

for the baryon spin $1/2$
\begin{equation}
	m[\Xi_{cc(2P)}^{++}]\approx m[\Xi_{cc(2P)}^+]=3885\pm 66 \mbox{ MeV,}
\end{equation}

for the baryon spin $3/2$
\begin{equation}
	m[\Xi_{cc(2P)}^{++}]\approx m[\Xi_{cc(2P)}^+]=4017\pm 66 \mbox{ MeV.}
\end{equation}

\section{Conclusion}
We have shown that the distortion of quark-gluon string inside the diquark structure of baryons can be numerically studied in the spectroscopy of doubly charmed baryons, since the corresponding positive corrections to the limit of local doubly heavy diquark are certainly valuable within the expected experimental accuracy. So, the masses of doubly charmed baryons in basic states are shifted upward. The additional impact  can be provided by comparison of mass observation for the basic states with the mass of quazi-stable excitation containing the scalar diquark.

It worth to mention, the non-locality of the diquark at distance $\lesssim 0.3$~fm can described within the form factor formalism as it was done in \cite{Ebert:2002ig}.   However, the form factor of diquark-gluon interaction is  not related to the  quark-gluon string distortion inside the diquark.
  
The clarification of quark-diquark structure in the case of doubly charmed baryons will serve for more reliable predictions of properties of exotic hardrons such as pentaquarks and tetraquarks \cite{Karliner:2016zzc,Ali:2017jda,Berezhnoy:2011xn}.

The production of doubly heavy baryons in hadron collisions  at energies of LHC has been calculated in \cite{Berezhnoy:1998aa,Chang:2006eu} within the framework of single parton scattering and color-singlet approximation. The contribution of double parton scattering is significant at such high energies, so it has been evaluated in \cite{Kiselev:2016rqj}. In addition, there are estimates in the model of intrinsic charm \cite{Koshkarev:2016rci,Koshkarev:2016acq}. 

The lifetimes and decay modes of doubly charmed baryons were calculated in the framework of operator product expansion \cite{Kiselev:1998sy,Guberina:1999mx}.

Supposing a systematic efficiency of LHC detectors to be sufficiently large for a lucky cascade mode of decay we expect that the doubly charmed baryons will be discovered in a short time\footnote{The SELEX result on the observation of doubly charmed baryons \cite{Mattson:2002vu,Ocherashvili:2004hi} has problems with the confirmation because of possible decay products misidentification and unexpectedly short lifetime \cite{Kiselev:2002an}.}\cite{Aaij:2013voa}, that provide us with the important data on both the spectroscopy and lifetimes of those baryons.

\acknowledgements This work is supported by Russian Foundation for Basic
Research, grant \#~15-02-03244. The work of V.V.K. is supported by Russian Ministry of Science and Education, project \#~3.9911.2017/BasePart.  A.V.B. acknowledges the support from MinES of RF (grant 14.610.21.0002, identification number RFMEFI61014X0002).

\bibliography{bib_cc}
\end{document}